\documentclass[english,aps,preprint]{revtex4}
\usepackage[T1]{fontenc}
\usepackage[latin9]{inputenc}
\usepackage{amsmath}
\usepackage{amssymb}
\usepackage{graphicx}

\makeatletter
\@ifundefined{textcolor}{}
{%
 \definecolor{BLACK}{gray}{0}
 \definecolor{WHITE}{gray}{1}
 \definecolor{RED}{rgb}{1,0,0}
 \definecolor{GREEN}{rgb}{0,1,0}
 \definecolor{BLUE}{rgb}{0,0,1}
 \definecolor{CYAN}{cmyk}{1,0,0,0}
 \definecolor{MAGENTA}{cmyk}{0,1,0,0}
 \definecolor{YELLOW}{cmyk}{0,0,1,0}
}

\makeatother

\usepackage{babel}
\begin{document}

\preprint{Brown-HET-1647}

\title{Event horizons and holography}

\author{Steven G. Avery}

\email{steven_avery@brown.edu}

\selectlanguage{english}%

\affiliation{Department of Physics, Brown University, Providence, RI 02912, USA}

\author{David A. Lowe}

\email{lowe@brown.edu}

\selectlanguage{english}%

\affiliation{Department of Physics, Brown University, Providence, RI 02912, USA}
\begin{abstract}
We consider the microcanonical ensemble of black holes in gravitational
theories in asymptotically anti-de Sitter spacetime with a conformal
field theory dual. We argue that typical black hole states show no
violations of general covariance on the horizon.
\end{abstract}
\maketitle

\section{introduction}

For black holes in asymptotically anti-de Sitter spacetime there are
two natural choices of vacua compatible with the symmetries. One such
vacuum is the analog of the Boulware vacuum \cite{Boulware:1974dm}:
positive frequency field modes far from the black hole, defined with
respect to the timelike Killing vector, annihilate the vacuum state.
Another natural choice is the analog of the Hartle--Hawking vacuum
\cite{Hartle:1976tp}, where positive frequency is defined with respect
to time translations of smooth global slices. This choice of vacuum
gives rise to entanglement between the left and right asymptotic regions
of the maximally extended AdS--Schwarzschild Penrose diagram.

Such geometric states can be viewed either in an entangled tensor-product
of two CFTs associated with the two asymptotic regions, or they may
be viewed as a density matrix in a single CFT. In the following, we
will restrict our attention to the second option. 

One immediately runs into an issue at the semiclassical level: a normalizable
scalar field mode in Schwarzschild anti-de Sitter contains both an
ingoing and an outgoing flux at the horizon. The outgoing flux produces
a divergent energy density as seen by an infalling observer crossing
the future horizon. At first sight, this seems to imply that the typical
finite energy states will be singular geometries, and moreover the
horizon will appear singular as it is approached from the outside.

Analyzing this problem from the point of view of the dual quantum
field theory in the microcanonical ensemble, Marolf--Polchinski (MP)
\cite{Marolf:2013dba} studied the number operator for Kruskal-like
modes, those natural from the viewpoint of a freely falling observer.
They argued in any eigenstate of such normalizable modes, this number
operator would always be of order 1. This then implies that typical
black holes always have violations of general covariance near the
horizon.

\section{Semi-classical approach\label{sec:Semi-classical-approach}}

\begin{figure}
\includegraphics[scale=0.5]{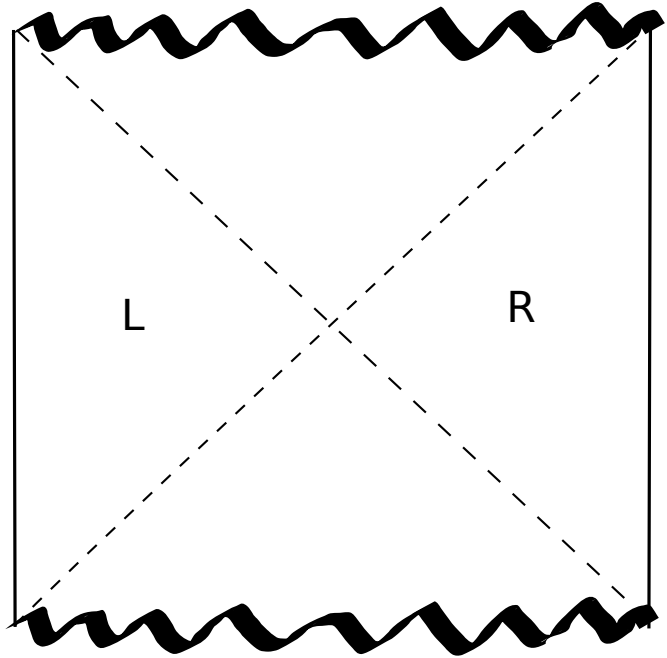}\caption{The Penrose diagram for Schwarzschild-anti de Sitter spacetime.}

\end{figure}
Let us re-examine this argument in more detail. The $b$-modes are
eigenstates of the timelike Killing vector at infinity, which is closely
associated with the CFT Hamiltonian. They argued that these modes
provide a complete basis for the energy eigenstates, along with a
set of possible other labels/modes. This is certainly correct in the
semiclassical limit, where one quantizes the modes around the black
hole geometry, ignoring issues of back-reaction. The Penrose diagram
for the anti-de Sitter Schwarzschild black hole is shown in figure
1.

A bulk scalar field in the right patch (R) may be decomposed as
\[
\phi=\sum_{k}b_{R,k}\phi_{R,k}+b_{R,k}^{\dagger}\phi_{R,k}^{*}
\]
where $k$ schematically represents the set of labels. We refer to
these modes as the $b$-modes, and they annihilate the Schwarzschild
vacuum state $b_{R,k}|0\rangle_{S}=0$. 

If the CFT description is to include a description of the black hole
interior, one must also consider a set of operators representing the
fields in the left patch (L), which propagate into the upper patch
in figure 1. Both sets of modes are needed to provide a complete description
of the field in the upper patch. The full decomposition, valid in
all coordinate patches is then
\[
\phi=\sum_{k}b_{L,k}\phi_{L,k}+b_{L}^{\dagger}\phi_{L,k}^{*}+b_{R,k}\phi_{R,k}+b_{R,k}^{\dagger}\phi_{R,k}^{*}
\]
and the Schwarzschild vacuum state is also annihilated by the left
$b$-modes $b_{L,k}|0\rangle_{S}=0$. Note that we define the mode
functions, $\phi_{L/R}$, in the above such that they only have support
in the appropriate (left/right) region.

One may also choose to decompose the field with respect to Kruskal
modes, which are analytic across the horizon
\[
\phi=\sum_{k}a_{k}\phi_{K,k}+a_{k}^{\dagger}\phi_{K,k}^{*}\,.
\]
We refer to these modes as the $a$-modes. These modes annihilate
the Kruskal vacuum $a_{k}|0\rangle_{K}=0$. These modes look rather
complicated when decomposed into frequencies with respect to the timelike
Killing vector at infinity. However as shown in \cite{Unruh:1976db},
these may be rewritten in terms of another set of operators $d_{L,k}$
and $d_{R,k}$ that annihilate $|0\rangle_{K}$ but are simply related
to the $b$-modes
\[
\phi=\sum_{k}\left(2\sinh\left(\beta\omega_{k}/2\right)\right)^{-1/2}\left(d_{R,k}\left(e^{\beta\omega_{k}/2}\phi_{R,k}+e^{-\beta\omega_{k}/2}\phi_{L,-k}^{*}\right)+d_{L,k}\left(e^{-\beta\omega_{k}/2}\phi_{R,-k}^{*}+e^{\beta\omega_{k}/2}\phi_{L,k}\right)\right)+\mathrm{h.c.}
\]
where $\beta$ is the inverse Hawking temperature of the black hole
and $\omega_{k}$ is the positive frequency associated with the mode
labeled by $k$. These operators are related to the $b$-mode operators
by a Bogoliubov transformation 
\begin{eqnarray*}
b_{L,k} & = & \left(2\sinh\left(\beta\omega_{k}/2\right)\right)^{-1/2}\left(e^{\beta\omega_{k}/2}d_{L,k}+e^{-\beta\omega_{k}/2}d_{R,-k}^{\dagger}\right)\\
b_{R,k} & = & \left(2\sinh\left(\beta\omega_{k}/2\right)\right)^{-1/2}\left(e^{\beta\omega_{k}/2}d_{R,k}+e^{-\beta\omega_{k}/2}d_{L,-k}^{\dagger}\right)
\end{eqnarray*}
which allows the different vacua to be related via
\begin{equation}
|0\rangle_{K}=\prod_{k}\exp\left(e^{-\beta\omega_{k}/2}b_{L,k}^{\dagger}b_{R,k}^{\dagger}\right)|0\rangle_{S}=\prod_{k}\sum_{n_{k}=0}^{\infty}e^{-\beta\omega_{k}n_{k}/2}|N_{b,L,k}=n_{k}\rangle\times|N_{b,R,k}=n_{k}\rangle\label{eq:hartlehawking}
\end{equation}
where $N_{b,L,k}$ and $N_{b,R,k}$ are the number operators for the
$b$-modes. It is important to keep in mind the energy in terms of
the $b$-mode eigenstates takes the form
\[
E=N_{b,L,k}\omega_{k}-N_{b,R,k}\omega_{k}
\]
so semiclassically any energy level has infinite degeneracy and the
entropy in the microcanonical ensemble diverges. Obviously this computation
ignores back-reaction, and should not be trusted. Likewise the finite
Bekenstein--Hawking entropy of the black hole cannot be obtained in
this kind of semiclassical quantum field theory in curved spacetime
approximation.

The number operator relevant for an infalling observer can be defined
as the number operator built from the Kruskal mode number operators
\begin{equation}
N_{d,k}=d_{L,k}^{\dagger}d_{L,k}+d_{R,k}^{\dagger}d_{R,k}\label{eq:infallnumber}
\end{equation}
and this annihilates $|0\rangle_{K}$. In fact, everything we say
applies to each term in the above separately. We will utilize this
expression momentarily.

\section{Euclidean quantum gravity approach}

This framework imposes periodicity in imaginary time to formulate
the canonical ensemble \cite{Hawking:1982dh}. The microcanonical
ensemble is then defined via a Laplace transform of the canonical
ensemble. In the gravitational sector, the correct Bekenstein--Hawking
black hole entropy is obtained, along with a contribution due to a
thermal bulk field modes. This contribution can be viewed as computing
the entropy of the reduced density matrix obtained by starting with
the pure state \eqref{eq:hartlehawking} and tracing over the left-modes.
For sufficiently large total energies, the microcanonical ensemble
is dominated by purely the black hole entropy contribution, with a
negligible term coming from the bulk modes.

\section{AdS/CFT Approach}

There is strong evidence the CFT is able to correctly reproduce the
Bekenstein--Hawking contribution to the entropy in the large mass
limit \cite{Gubser:1996de}. The entropy is reproduced up to an overall
constant that is difficult to determine precisely, because the CFT
is strongly coupled in the limit that it is dual to a gravitational
phase.

This approach must also yield significant corrections to the approach
of section \ref{sec:Semi-classical-approach}. Let us focus on the
case of the four-dimensional bulk spacetime theory for the sake of
definiteness. The boundary of the theory is $S^{2}\times\mathbb{R}$,
with the $\mathbb{R}$ factor corresponding to the time coordinate.
Because the spatial sections are compact spheres, the energy spectrum
of the conformal field theory becomes discrete. This induces a particular
cutoff on the spectrum of the bulk theory.

The argument of MP \cite{Marolf:2013dba} proceeds by assuming that
the $b_{R}$-mode number eigenstates %
\footnote{In fact, they utilize a regularized version of the $b$-modes that
are smoothened near the horizon. This smoothening does not affect
the arguments we present here.%
} provide approximate energy eigenstates in the exact theory dual to
the CFT. Presumably they are also assuming the states also carry $b_{L}$-mode
quantum numbers, as well as other possible labels. The exact bulk
spectrum should be discretized in some way, to match that of the CFT.
Within any such a number eigenstate, the expectation value of \eqref{eq:infallnumber}
is greater than or equal to 1. This leads to a bound on the average
of $N_{d}$ in the microcanonical ensemble that is also greater than
or equal to 1. One then concludes that general covariance is violated.
Moreover, each $b$-mode number eigenstate leads to a divergent stress
energy tensor outside the horizon, as described in \cite{Lowe:2013zxa}. 

However the discussion of section \ref{sec:Semi-classical-approach}
shows there is a counting problem, once one tries to diagonalize the
$N_{d}$ operator using $b_{R}$-mode eigenstates. There are far more
semiclassical modes than there are exact quantum states. It is therefore
wishful thinking that the analog of the $N_{d}$ operator in the exact
theory can be approximately diagonalized in terms of such eigenstates.
The typicality argument of MP for firewall states therefore breaks
down.

Therefore if one wants to take seriously the matching between the
finite entropy of the microcanonical ensemble $S=S(M)$ and the finite
entropy of the CFT, then the Hilbert space of the exact quantum description
must be much smaller than the infinite dimensional Hilbert subspace
one has in the semiclassical description, at fixed total energy. 

Since this is ultimately determined by the properties of a strongly
coupled conformal field theory and the bulk--boundary dictionary,
for the moment we are free to postulate what properties the exact
modes should have. In particular, we expect if a black hole is formed
by sending matter in from the boundary of empty anti-de Sitter spacetime,
that the bulk field states should be determined by following unitary
evolution from the vacuum up into the black hole region. This implies
a correlation between particular $b_{R}$-modes and $b_{L}$-modes.
We further postulate that a good approximation to the energy eigenstates
are actually the $a$-modes, equipped with a cutoff. In the presence
of this cutoff, one cannot simply rediagonalize and use the $b$-mode
number eigenstates. Expressed in terms of the $b$-modes, only very
particular complex superpositions correspond to exterior bulk states
in the exact theory, such as the vacuum state \eqref{eq:hartlehawking},
and states obtained by acting with the operators $d_{L}^{\dagger}$
and $d_{R}^{\dagger}$.

It should be noted, that while the $d_{R}^{\dagger}$ operators contain
a component inside the horizon of the black hole, this contribution
is suppressed by a factor $e^{-\beta\omega_{k}}$ relative to the
component outside. This is sufficient for these modes to be used to
define a local effective field theory outside the horizon, with some
proper distance cutoff scale in the bulk close to the Planck length. 

Another argument in favor of using the $d_{L}^{\dagger}$ and $d_{R}^{\dagger}$
modes comes from the lattice black hole analysis of \cite{Corley:1997ef}.
There it was found that a short distance regulator appropriate for
a freely falling frame, does not commute with the Hamiltonian associated
with the timelike Killing vector. Thus $b$-mode eigenstates are not
preserved under time evolution. The $d$-modes avoid this issue, by
including entanglement between the interior and exterior. Tracing
over the interior component then leads to an approximately thermal
density matrix for the $b$-modes, which is indeed preserved under
time evolution.

\section{Conclusions}

We have argued that the result of MP \cite{Marolf:2013dba}, that
a typical black hole in a theory with a dual gauge theory sees a violation
of general covariance at the horizon, relies on expanding the Hilbert
space well-beyond that of the exact theory. Because the exact states
will only be close to very special states in this expanded Hilbert
space, the typicality statement fails. We have given examples of sequences
of states in the microcanonical ensemble where general covariance
is expected to hold near the horizon, and conjecture that such states
provide a good approximation to the exact states of the quantum theory.
\begin{acknowledgments}
D.L. thanks Larus Thorlacius for discussions. This research was supported
in part by DOE grant DE-SC0010010 and an FQXi grant.
\end{acknowledgments}
\bibliographystyle{apsrev}
\bibliography{firewall}

\end{document}